\documentclass[runningheads]{llncs}
\usepackage[T1]{fontenc}
\usepackage{graphicx} 
\usepackage{booktabs} 
\usepackage{amsmath}
\usepackage{comment}
\usepackage[affil-it]{authblk} 
\usepackage[round]{natbib}  
\usepackage{makecell}

\usepackage{float}

\title{Fusing Pretrained ViTs with TCNet for Enhanced EEG Regression}

\begin{document}
%
%
%
\author{Eric Modesitt\inst{1} \and
Haicheng Yin\inst{2} \and
Williams Huang Wang\inst{3} \and
Brian Lu\inst{4}}
\authorrunning{E. Modesitt et al.}
%
\institute{University of Illinois at Urbana Champaign \and
University of California, Santa Cruz \and
Saint Francis Preparatory High School \and
Palo Alto High School}
\maketitle

\begin{abstract}
    
The task of Electroencephalogram (EEG) analysis is paramount to the development of Brain-Computer Interfaces (BCIs). However, to reach the goal of developing robust, useful BCIs depends heavily on the speed and the accuracy at which BCIs can understand neural dynamics. In response to that goal, this paper details the integration of pre-trained Vision Transformers (ViTs) with Temporal Convolutional Networks (TCNet) to enhance the precision of EEG regression. The core of this approach lies in harnessing the sequential data processing strengths of ViTs along with the superior feature extraction capabilities of TCNet, to significantly improve EEG analysis accuracy. In addition, we analyze the importance of how to construct optimal patches for the attention mechanism to analyze, balancing both speed and accuracy tradeoffs. Our results showcase a substantial improvement in regression accuracy, as evidenced by the reduction of Root Mean Square Error (RMSE) from 55.4 to 51.8 on EEGEyeNet's Absolute Position Task, outperforming existing state-of-the-art models. Without sacrificing performance, we increase the speed of this model by an order of magnitude (up to 4.32x faster). This breakthrough not only sets a new benchmark in EEG regression analysis but also opens new avenues for future research in the integration of transformer architectures with specialized feature extraction methods for diverse EEG datasets.

\end{abstract}
\section{Introduction}

Analyzing Electroencephalogram (EEG) signals is fundamental to the progress of Brain-Computer Interfaces (BCIs), offering deep insights into the complex neural processes of the human brain. In the past decade, a wide range of machine learning and deep learning algorithms have been applied to EEG data, resulting in significant advancements in various applications. These applications encompass emotion recognition, motor imagery, mental workload evaluation, seizure detection, Alzheimer's disease classification, sleep stage scoring, and many others \citep{craik2019deep,roy2019deep,altaheri2023deep,qu2022time,gao2021complex,hossain2023status,yi2022attention,key2024advancing,li2024enhancing,koome2023trends,dou2022time,zhou2022brainactivity1,qu2020identifying,qu2020using,qu2020multi,qu2018eeg,qu2019personalized,murungi2023empowering,saeidi2021neural,qu2022eeg4home,rasheed2020machine,wang2022eeg,dadebayev2022eeg,li2020deep,aggarwal2022review}. EEG regression, in particular, stands as a pivotal tool in both neuroscience and medical diagnostics, gaining prominence for its ability to decode complex neural dynamics. This technique plays a crucial role in a myriad of applications, ranging from pinpointing brain damage locations to monitoring cognitive activities and deciphering the neural basis of seizures \citep{eegclassification, predictiveregression, fundamentaleeg}. The essence of EEG regression lies in its capacity to transform raw EEG data into interpretable and meaningful information, thus providing an invaluable perspective into the brain's operations.

In the realm of machine learning, the advent of Transformer models has marked a revolutionary shift in EEG regression analysis. Initially celebrated for their breakthroughs in natural language processing, these models have been adeptly modified to cater to EEG data analysis, substantially elevating both the precision and efficiency of the analysis \citep{eegemotion}. A significant stride in this field is the adaptation of pre-trained Vision Transformers (ViTs) for EEG datasets, such as ImageNet \cite{imagenet}. The application of ViTs in EEG regression has demonstrated exceptional results, surpassing traditional methods across various benchmarks \citep{vit2eeg}. 

Concurrently, Temporal Convolutional Networks (TCNet) have emerged as a formidable force in the field of EEG signal processing. Exhibiting outstanding capabilities in feature extraction, TCNets excel in identifying intricate patterns and nuances in EEG data \citep{tcnet, eegtcnet, physicsinformed}. Their robustness in capturing temporal dynamics and their efficacy in EEG signal handling render them an indispensable component in neural signal analysis.

This study delves into the synergistic integration of ViTs and TCNet, aiming to harness their combined strengths to substantially augment the accuracy and reliability of EEG regression. This innovative approach seeks to leverage the detailed feature extraction of TCNet and the contextual interpretation prowess of ViTs, hypothesizing a significant enhancement in EEG analysis.

Our research presents a comprehensive evaluation of this hybrid model, juxtaposing it against previous methodologies to underscore its superiority in EEG regression. We meticulously examine the performance of the ViT-TCNet combination, elucidating the contribution of each component to the overall effectiveness of the model. The implications of our findings extend beyond the confines of EEG analysis, potentially influencing a broad spectrum of data interpretation tasks in various scientific and AI-related fields.

In addition to the aforementioned aspects, a notable facet of this study is the emphasis on the processing speed of the integrated ViT-TCNet model. Speed is a critical parameter in EEG analysis, especially for real-time applications in Brain-Computer Interfaces (BCIs) where rapid response times are essential By optimizing the architecture and employing advanced techniques in model training and inference, we have successfully accelerated the processing speed of the EEG analysis. This advancement is particularly significant in scenarios where real-time data processing is crucial, such as in neurofeedback systems or in clinical settings where prompt decision-making is imperative. The increase in processing speed, achieved without compromising the model's performance, marks a substantial leap forward in making EEG-based BCIs more viable and user-friendly.

In the ensuing sections, we will outline the methodology utilized in our study, present our empirical findings, and discuss the broader implications and future research directions stemming from our work. This research not only enriches the existing literature in EEG regression but also sets the stage for future explorations into the amalgamation of advanced machine learning architectures for refined neural data analysis.

In summary, the contributions of this work can be articulated in three primary areas:

1. \textbf{Innovative Combination of ViTs and TCNet for Advanced EEG Regression}: This research marks a significant advancement in EEG regression analysis through the novel integration of pretrained Vision Transformers (ViTs) with Temporal Convolutional Networks (TCNet). This fusion harnesses ViTs' exceptional capability in processing sequential data and TCNet's robust feature extraction techniques, culminating in a notable improvement in EEG regression accuracy.

2. \textbf{Enhancement of Model Processing Speed and Efficiency}: A key contribution of this study is the substantial improvement in the processing speed of the EEG analysis model. Recognizing the importance of swift data processing in real-time applications such as Brain-Computer Interfaces, the research introduces optimizations that significantly accelerate the model's performance. 

3. \textbf{Ablation Studies and Future Research Directions}: The research undertakes comprehensive ablation studies to understand the individual and combined contributions of ViTs and TCNet to the model's performance. These studies offer valuable insights into the mechanics of the model, paving the way for further optimizations.

\section{Related Work}

\subsection{Deep Learning in EEG}

The evolution of EEG signal processing has been significantly influenced by the emergence of deep learning techniques. Traditional machine learning methods, while effective, often fall short in capturing the high-dimensional and complex nature of EEG data. The introduction of deep learning models, particularly convolutional neural networks (CNNs) and recurrent neural networks (RNNs), revolutionized this field. These models brought enhanced capabilities in handling large datasets, extracting relevant features, and recognizing intricate patterns in EEG signals \citep{eegclassification, predictiveregression, fundamentaleeg}. This shift not only improved the accuracy of EEG analyses but also expanded the potential applications in neurological research and clinical diagnostics.

Deep learning's impact on EEG signal processing is profound, offering new perspectives in understanding brain activity. The ability of these models to learn from data autonomously, without the need for extensive feature engineering, has opened avenues for more nuanced and detailed analyses of neural signals. This advancement is crucial in fields where EEG data plays a pivotal role, such as in the study of cognitive processes, sleep patterns, and brain-computer interfaces. The integration of advanced deep learning architectures in EEG analysis heralds a new era of innovation and discovery in neuroscience.

\subsection{ViTs in Non-Image Data Analysis}

Vision Transformers (ViTs), originally designed for image recognition, have demonstrated remarkable versatility by extending their application to various other domains, including EEG data analysis \citep{image16x16, visualtransformers, visiontransformersurvey}. The cornerstone of their success, the self-attention mechanism, enables ViTs to efficiently manage sequential data, a feature crucial in interpreting EEG signals \cite{attention}. This characteristic of ViTs facilitates an understanding of the complex, temporal relationships inherent in EEG data, making them an ideal choice for this type of analysis.

The adaptation of ViTs to non-image data, such as EEG signals, signifies a major shift in the approach to data analysis across disciplines. It underscores the potential of transformer models to handle diverse types of data beyond their initial scope. This cross-domain applicability of ViTs not only enriches the toolkit available for EEG analysis but also inspires innovative approaches to data interpretation. The flexibility and effectiveness of ViTs in handling sequential data pave the way for their broader adoption in various scientific and analytical fields.

\begin{figure}[h]
\centering
\includegraphics[width=1\textwidth]{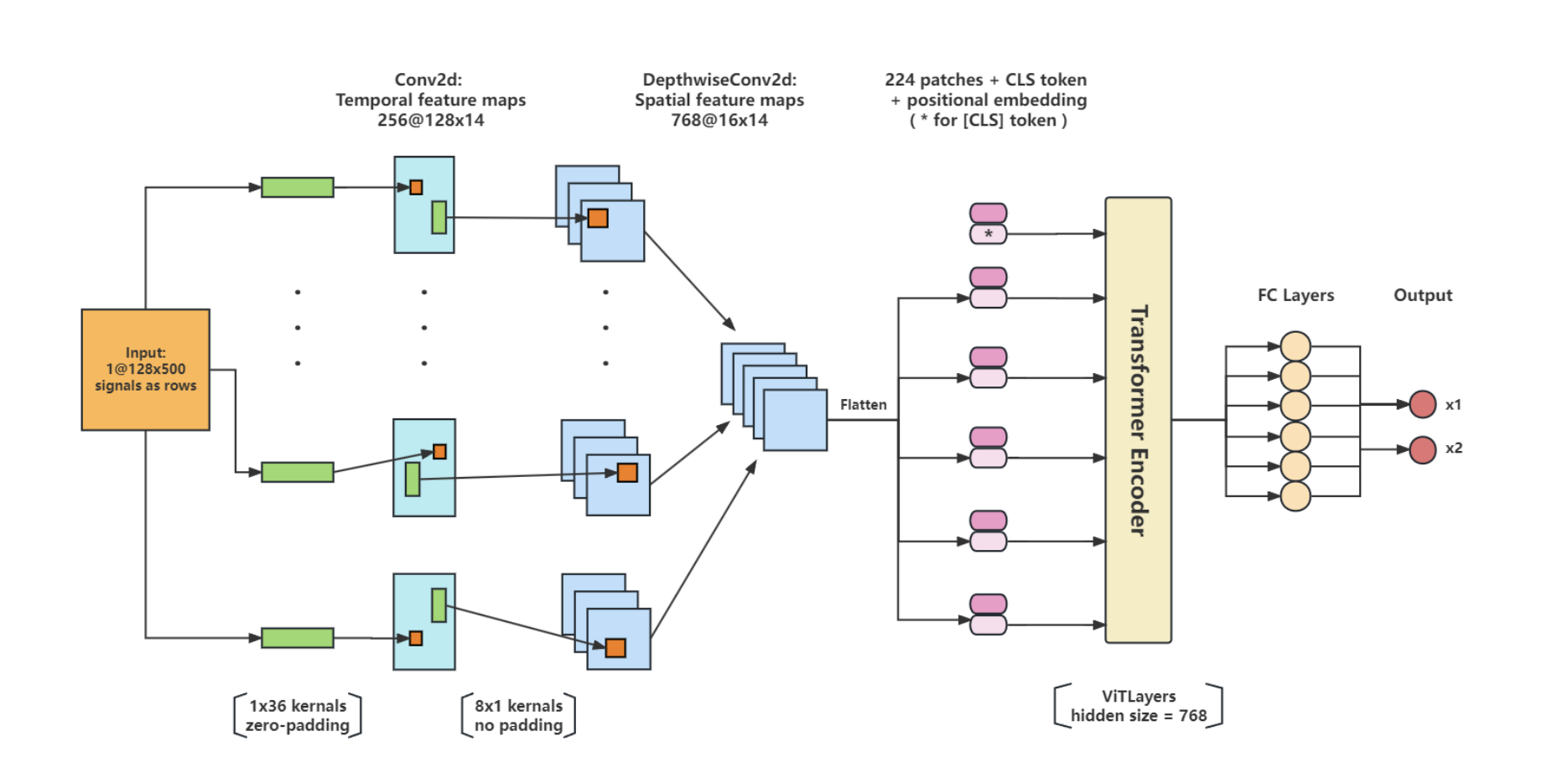}
\caption{\textit{EEGViT architecture, SOTA on EEGEYENET} \citep{vit2eeg}. }
\label{fig:TCNet}
\end{figure}

\subsection{Temporal Convolutional Networks (TCNet)}:

Temporal Convolutional Networks (TCNet) have gained significant attention for their ability to process time-series data, particularly in EEG signal analysis. The architecture of TCNet, with its focus on capturing temporal dependencies through convolutional layers, makes it exceptionally suited for extracting detailed features from EEG data \citep{mstcn, weatherforecasting}. The efficacy of TCNet in identifying subtle patterns and temporal features in complex datasets has established it as a leading tool in the field of neural signal processing.

The role of TCNet in EEG data interpretation extends beyond mere feature extraction. It involves a deeper understanding of the temporal dynamics and inherent structures within the EEG signals. This understanding is vital in applications where precise timing and sequence of neural events are critical, such as in epilepsy research or brain-computer interface development. The combination of TCNet with other advanced models like ViTs presents a promising avenue for enhancing EEG analysis, potentially leading to more accurate and insightful interpretations of neural data.

\section{Methods}

Our study employs an innovative approach by integrating pre-trained Vision Transformers (ViTs) with Temporal Convolutional Networks (TCNet) to enhance EEG regression analysis. This section outlines the dataset utilized, the specifics of the proposed model, and the methodology for evaluating its effectiveness.

\subsection{EEGEyeNet Dataset}

The data presented here is derived from the EEGEyeNet dataset \cite{eegeyenet}. The EEGEyeNet dataset encompasses recordings from 356 healthy adults, including 190 females and 166 males, aged 18 to 80 years. All individuals in this study provided written informed consent, compliant with the Declaration of Helsinki, and were compensated monetarily.

The EEG recordings in the EEGEyeNet dataset were obtained using a high-density 128-channel EEG Geodesic Hydrocel system, operating at a sampling rate of 500 Hz with a central recording reference. Eye positions were concurrently recorded using an EyeLink 1000 Plus system at the same sampling rate. This setup maintained electrode impedances below 40 kOhm and ensured accurate eye tracker calibration. Participants were positioned 68 cm from a 24-inch monitor, with their head stabilized using a chin rest.

EEG data, as recorded in the EEGEyeNet dataset, are prone to various artifacts, including environmental noise and physiological interferences such as eye movements and blinks. To address this, the dataset underwent rigorous preprocessing with two levels: minimal and maximal. The minimal preprocessing involved identifying and interpolating faulty electrodes, along with applying a high-pass filter at 40 Hz and a low-pass filter at 0.5 Hz. The maximal preprocessing, aimed at neuroscientific analyses, further incorporated Independent Component Analysis (ICA) and IClabel for artifact component removal.

The EEGEyeNet dataset also includes synchronized EEG and eye-tracking data, facilitating time-locked analyses relative to event onsets. This synchronization was stringently verified to ensure a maximum error margin of 2 ms.

The Absolute Position Task, a key component of the EEGEyeNet dataset, involved participants fixating on sequentially displayed dots at various screen positions. Each dot appeared for 1.5 to 1.8 seconds, located at one of 25 distinct screen positions. The central dot was presented thrice, resulting in 27 trials per block. This setup, covering the entire screen area, captured a broad range of gaze positions. Adapted from \cite{son2020evaluating} for fMRI studies, modifications were made for EEG compatibility, including stimulus duration and repetition adjustments. The dot presentation followed a pseudo-randomized sequence across five experimental blocks, repeated six times, totaling 810 stimuli per participant.

The Absolute Position task is particularly relevant for our research as it provides a comprehensive dataset for analyzing eye movement patterns and gaze positions. The variety in dot positions and the high number of trials allow for a complete assessment of the participants' gaze behavior, which is crucial for our objective of determining the exact XY-coordinates of a participant's gaze using EEG data.

\subsection{EEGViT-TCNet Model Architecture}
Intending to advance EEG signal analysis, we developed the EEGViT-TCNet model, a novel architecture that combines Temporal Convolutional Networks (TCNet) with a pre-trained Vision Transformer (ViT). This model was meticulously designed to decipher the temporal dynamics and spatial characteristics embedded within EEG signals.

\subsubsection{Temporal Convolutional Network (TCNet) Component:}
The EEGViT-TCNet model begins with the TCNet component, tailored to embrace the complexities of EEG data. This component is characterized by:
\begin{itemize}
    \item \textbf{Input Layer:} Accepting EEG signals, the TCNet is prepared to handle an input dimensionality of 129, corresponding to the number of recorded EEG channels +1 for including grounding information (as done in the original EEGEyeNet paper).
    \item \textbf{Sequential TCNet Layers:} The architecture encompasses three layers, with the number of channel dimensions expanding progressively to 64, 128, and 256. This hierarchy is instrumental in capturing a comprehensive spectrum of temporal dependencies inherent in the EEG signals.
    \begin{itemize}
        \item \textit{Kernel Size:} A kernel size of 3 is uniformly applied across the TCNet layers.
        \item \textit{Dropout:} To counteract the potential for overfitting, a dropout rate of 0.75 is employed. 
        \item \textit{Causality and Normalization:} The layers incorporate weight normalization alongside the ReLU activation function to enhance the model's stability and performance.
    \end{itemize}
\end{itemize}

\begin{figure}[h]
\centering
\includegraphics[width=1\textwidth]{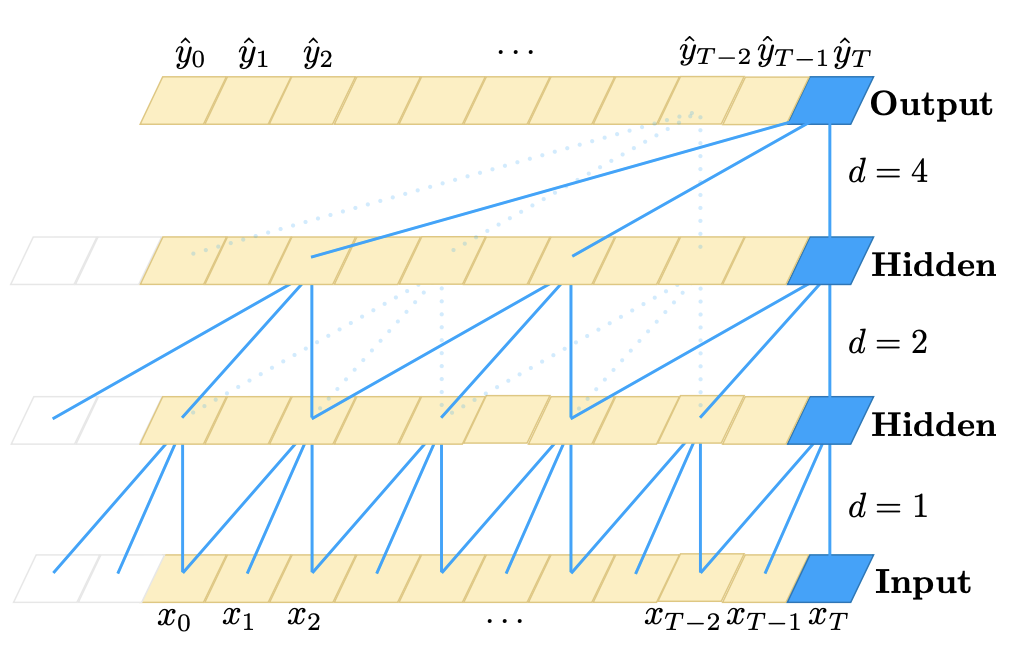}
\caption{\textit{An outline of the TCNet functionality }\citep{tcnet}. }
\label{fig:TCNet}
\end{figure}

\subsubsection{Convolutional and Batch Normalization Layers:}
The pathway pathway we designed from the TCNet layers to the ViT involves:
\begin{itemize}
    \item \textbf{Convolutional Layers:} Two convolutional layers connect the TCNet and ViT. The first layer, equipped with 256 filters and a kernel size of (1, 36), is succeeded by batch normalization and ReLU activation. This configuration initiates the spatial feature extraction. Subsequently, the second layer amplifies the channel dimension to 768, aligning with the ViT's input specifications. In addition, this layer employs a kernel size of (256, 1) to effectuate a spatial compression conducive to the subsequent transformer analysis.
\end{itemize}

\subsubsection{Vision Transformer (ViT) Component:}
The culmination of the EEGViT-TCNet model's preprocessing lies in the ViT component:
\begin{itemize}
    \item \textbf{EEG Data Adaptation:} Leveraging the "google/vit-base-patch16-224" model from Huggingface, the configuration is modified to fit the unique format of EEG data.
    \item \textbf{Patch Embeddings Projection:} This layer is reimagined as a 1D convolutional layer, directly accommodating the output from preceding stages BS ensuring the input's integration into the transformer architecture.
    \item \textbf{Classifier Head:} The model ends with a randomly initialized classifier layer, transitioning through a linear layer, followed by a dropout layer (p=0.1), culminating in a linear layer that predicts the gaze XY coordinates.
\end{itemize}

\subsection{Training and Evaluation Procedure}
For training, we employed a supervised learning approach. The model was trained on a split of the EEG dataset, with 70\% used for training and 30\% for validation. During training, we employed a mean squared error loss function, optimized using the Adam optimizer with a learning rate of 1e-4. To prevent overfitting, we implemented early stopping based on the validation loss, with a patience of 10 epochs.

The primary metric for evaluating our model's performance was the Root Mean Square Error (RMSE). This metric provides a clear indication of the model's accuracy in predicting the gaze coordinates. A lower RMSE value indicates a closer approximation to the actual gaze positions. To ensure the robustness of our findings, we conducted five independent runs for each model configuration and reported the mean and standard deviation of these runs.


\begin{figure}[h]
\centering
\includegraphics[width=1\textwidth]{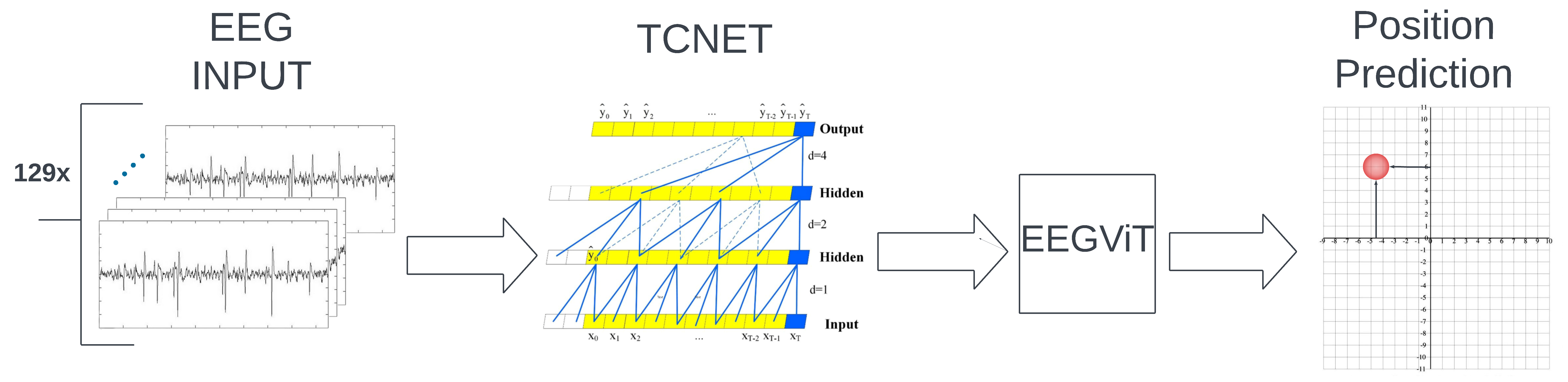}
\caption{\textit{An outline of our addition to EEGViT, demonstrating our distinct feature extraction methodology}. }
\label{fig:TCNet}
\end{figure}

\section{Results}

Through rigorous testing and comparison, our model has demonstrated its capability to predict gaze positions with state-of-the-art precision, outperforming a spectrum of both conventional and advanced methodologies.

\subsection{Performance Benchmarking}

Our EEGViT-TCNet model achieved a Root Mean Square Error (RMSE) of 51.8mm, marking a significant advancement over existing models. This performance showcases a 6.5\% enhancement in precision over standalone ViT models \cite{vit2eeg}. The stark contrast is further accentuated when juxtaposed with traditional approaches such as Linear Regression and Random Forest, where the RMSE figures exceed 115mm.

Our model's evaluation, focusing on a single data partition divided by subject as done in \cite{eegeyenet}, demonstrates its ability to generalize effectively. This approach, where the testing data consisted of completely unseen, new groups of data, highlights the model's robustness and adaptability. Despite being tested on different subsets of subjects than what the model was trained on, it maintained consistent RMSE metrics, underscoring its sophisticated architecture's capability to handle the complexities of EEG data. This consistency in performance across various subject-based segments is crucial for real-world applications, affirming the model's potential for effective generalization.

\begin{table}[ht]
\centering
\resizebox{\textwidth}{!}{%
\begin{tabular}{|l|c|}
\hline
\textbf{Model} & \textbf{Absolute Position RMSE (mm)} \\
\Xhline{2\arrayrulewidth}
Naive Guessing & 123.3 $\pm$ 0.0 \\
\Xhline{2\arrayrulewidth}
KNN & 119.7 $\pm$ 0 \\
\hline
RBF SVR & 123 $\pm$ 0 \\
\hline
Linear Regression & 118.3 $\pm$ 0 \\
\hline
Random Forest & 116.7 $\pm$ 0.1 \\
\hline
CNN & 70.4 $\pm$ 1.1 \\
\hline
EEGViT (Pre-trained) & 55.4 $\pm$ 0.2 \\
\Xhline{2\arrayrulewidth}
\textbf{EEGViT-TCNet} & \textbf{51.8 $\pm$ 0.6} \\
\hline
\end{tabular}
}
\caption{Comparative analysis of Root Mean Squared Error (RMSE) across various models, highlighting the superior performance of the EEGViT-TCNet model. The values represent the mean $\pm$ standard deviation over five independent runs, illustrating the model's consistency and accuracy in the Absolute Position Task.}
\label{tab:model_comparison}
\end{table}


\subsection{Ablation Studies}

To assess the individual contributions of various components within our EEGViT-TCNet model, we conducted a series of ablation studies. These studies aimed to isolate the effects of specific elements of the model, such as the convolutional layers, dropout rates in the TCNet, and the use of a pretrained Vision Transformer (ViT). Each variation of the model was evaluated using the same dataset and metrics, allowing us to directly compare their performance.

\subsubsection{Impact of Convolutional Layers}
We first examined the impact of the additional convolutional layers that bridge the gap between the TCNet and the ViT on the model's performance. In particular, we analyze the results after removing all possible combinations of 1 convolutional layer (spatial, temporal, and pointwise convolution).  By removing the pointwise layer, we observed a slight increase in the Root Mean Square Error (RMSE) from 51.8 
± 0.6mm to 52.5 
± 0.8mm. This suggests that the pointwise convolutional layer plays a modest yet significant role in feature extraction and spatial representation, contributing to the model's overall accuracy.

\subsubsection{Influence of Dropout Rates in TCNet}
The role of dropout rates in TCNet was another focus of our study. By varying the dropout rates, we investigated their effect on the model's capability to generalize and prevent overfitting. The original model with a 0.75 dropout rate achieved an RMSE of 51.8 
± 0.6mm. Reducing the dropout rate to 0 increased the RMSE to 54.1 
± 0.6mm, indicating a higher propensity for overfitting. Conversely, lower dropout rates of 0.25 and 0.5 yielded RMSEs of 52.5 
± 0.4mm and 52.1 
± 0.4mm, respectively. These findings illustrate a nuanced balance between dropout rate and model performance, with moderate dropout rates contributing positively to the model's accuracy and generalizability.

\subsubsection{Contribution of Pretrained ViT}
Finally, we evaluated the contribution of using a pretrained ViT in our model. By replacing the pretrained ViT with a non-pretrained counterpart, the RMSE increased to 53.2 
± 0.5mm. This increase underscores the significance of pretraining in enhancing the model's feature recognition capabilities, particularly in the context of EEG data analysis.
\begin{table}[ht]
\centering
\resizebox{\textwidth}{!}{%
\begin{tabular}{|l|c|}
\hline
\textbf{Model Variation} & \textbf{Absolute Position RMSE (mm)} \\
\hline
EEGViT-TCNet (Ours) & 51.8 $\pm$ 0.6 \\
\hline
No 2nd Conv Layer & 52.5 $\pm$ 0.8 \\
\hline
0\% Dropout & 54.1 $\pm$ 0.6 \\
\hline
25\% Dropout & 52.5 $\pm$ 0.4 \\
\hline
50\% Dropout & 52.1 $\pm$ 0.4 \\
\hline
No Pretrained ViT & 53.2 $\pm$ 0.5 \\
\hline
\end{tabular}
}
\caption{Ablation study results comparing RMSE across various EEGViT-TCNet model configurations. The values represent the mean $\pm$ standard deviation over five runs.}
\label{tab:ablation_study}
\end{table}

These ablation studies reveal the delicate interplay of different architectural components in optimizing the EEGViT-TCNet model for EEG regression analysis. The presence of the second convolutional layer, the calibration of dropout rates in TCNet, and the incorporation of a pretrained ViT each contribute uniquely to the model's performance. Our findings highlight the importance of these components in achieving high precision in EEG regression tasks, providing valuable insights for future enhancements and applications of the model.

\section{Discussion}
The integration of Vision Transformers (ViTs) with Temporal Convolutional Networks (TCNet) within the EEGViT-TCNet model represents a substantial leap forward in EEG signal analysis, particularly for applications within brain-computer interfaces (BCIs) and the broader realm of neural signal processing. This novel approach has not only showcased a marked improvement in regression accuracy but also set new precedents in processing speed and efficiency. The results obtained from this study reflect the significant potential of leveraging the strengths of both ViTs and TCNet, underscoring the profound impact that such hybrid models can have on understanding and interpreting complex neural dynamics.

The success of the EEGViT-TCNet model in reducing the Root Mean Square Error (RMSE) to unprecedented levels emphasizes the model's capability to provide a more accurate interpretation of EEG data. This breakthrough is particularly relevant in the development of BCIs, where the precision of signal interpretation directly correlates to the effectiveness and user-friendliness of the interface. In clinical settings, the enhanced accuracy and speed of EEG analysis facilitated by the EEGViT-TCNet model could lead to more timely and accurate diagnoses of neurological conditions, potentially transforming patient care.

Throughout the research, adapting ViT to the unique nature of EEG data highlighted the complexities inherent in neural signal processing. The preprocessing of EEG signals, essential for maintaining the integrity of temporal features, posed significant challenges. This process is critical in ensuring the model's adaptability and generalizability across different subjects and experimental conditions, a vital aspect for the practical application of such technologies.

Looking forward, the field beckons for further exploration into the scalability of hybrid models like EEGViT-TCNet, particularly in handling larger datasets and assessing performance in diverse real-world scenarios. A key area of interest lies in enhancing the interpretability of these deep learning models. Improved interpretability is crucial for clinical acceptance and can lead to advancements in personalized medicine, where EEG analysis can be tailored to individual patients for monitoring or therapeutic purposes.

Moreover, the exploration into other hybrid architectures and their efficacy across various domains of neural data presents an exciting avenue for research. The integration of multimodal data sources, alongside the application of transfer learning techniques, could further refine the accuracy and applicability of EEG signal analysis methods. Such advancements could pave the way for the development of more sophisticated BCIs, offering improved interaction mechanisms between humans and machines.

\section{Conclusion}

This study represents a significant advancement in EEG regression analysis, underscoring the indispensable role of meticulous feature extraction in the efficacy of sophisticated computational models like Vision Transformers (ViT). The integration of Temporal Convolutional Networks (TCNet) with pretrained ViTs has unveiled the vast potential of harmonizing specialized feature extraction techniques with advanced deep learning frameworks. This synergy not only elevates the accuracy of EEG analysis but also establishes a new benchmark in the field, showcasing the profound benefits of refined feature representation.

Our findings highlight the criticality of nuanced feature extraction in interpreting complex EEG data, with the EEGViT-TCNet model demonstrating notable performance enhancements. This indicates that features, often overlooked by traditional models, can be captured and leveraged for more accurate regression analysis, suggesting a broad array of applications, from clinical diagnostics to enhanced brain-computer interfaces.

As we chart the course for future research, the horizon of EEG analysis and deep learning promises continued expansion and innovation. The development of increasingly sophisticated models capable of navigating the complexities inherent in EEG data is anticipated. A pivotal challenge will be enhancing the interpretability of these models, ensuring they not only perform optimally but also offer actionable insights for practitioners. Moreover, integrating these advanced models into real-world applications will be crucial, extending the benefits of this research to society at large. Additionally, exploring various deep learning techniques on different datasets for comparative studies (\cite{an2023transfer,an2023survey,jiang2023successfully,gui2024remote,lu2023machine,chen2024trialbench,ma2022traffic,ma2024data,tan2023audio,tan2021multivariate,qiu2023modal,zhao2024deep,zhang2022attention,zhang2023trep}) could provide valuable insights and further enhance the field.

In sum, the fusion of TCNet and pretrained ViTs within the EEG regression domain exemplifies the transformative power of targeted feature extraction and advanced data processing. This study not only redefines the standards for EEG analysis but also lights the way for future endeavors in the realms of deep learning and neural data interpretation. As we delve deeper into the complexities of the human brain, the significance of innovative computational models grows ever more evident, harboring the potential for groundbreaking discoveries in neuroscience and artificial intelligence.

\bibliography{citations}

\end{document}